\documentclass[12pt]{iopart}

\usepackage{epsfig}

\begin{document}

%\draft

\title
{\sf Noble gas films on a decagonal AlNiCo quasicrystal}

\author{
W Setyawan$^1$, R D Diehl$^2$, N Ferralis$^3$, M W Cole$^2$ and S Curtarolo$^1$}
\address{
$^1$ Department of Mechanical Engineering and Materials Science, Duke University, Durham, NC 27708 \\
$^2$ Department of Physics and Materials Research Institute, Penn State University, University Park, PA 16801\\
$^3$ Department of Chemical Engineering, University of California, Berkeley, CA 94720\\
}
\ead{stefano@duke.edu}

\date{\today}

\begin{abstract}
Thermodynamic properties of Ne, Ar, Kr, and Xe adsorbed on an Al-Ni-Co
quasicrystalline surface (QC) are studied with Grand Canonical Monte
Carlo by employing Lennard-Jones interactions with parameter values
derived from experiments and traditional combining rules. In all the
gas/QC systems, a layer-by-layer film growth is observed at low
temperature. The monolayers have regular epitaxial fivefold
arrangements which evolve toward sixfold close-packed structures as
the pressure is increased.  The final states can contain either considerable
or negligible amounts of defects.  
In the latter case, there occurs a structural transition from
five to sixfold symmetry which can be described by introducing an order parameter,
whose evolution
characterizes the transition to be continuous or discontinuous as in
the case of Xe/QC (first-order transition with associated latent
heat).  By simulating fictitious noble gases, we find that the
existence of the transition is correlated with the size mismatch
between adsorbate and substrate's characteristic lengths. A simple
rule is proposed to predict the phenomenon.
\end{abstract}

\section{Introduction}
\label{section_introduction}

A tremendous interest in surface structures and phase transitions grew
out of the fact that surface systems are ideal for exploring the
effects of competing interactions \cite{schickreview,gcmc3}. A form of
competing interactions seen in adsorption involves either a length
scale or a symmetry mismatch between the adsorbate-adsorbate
interaction and the adsorbate-substrate interaction
\cite{sander1984,fairobent1982}.  Some consequences of such mismatches
include density modulations \cite{modulationAg1,modulationAg2}, domain
walls \cite{domainwallXe}, epitaxial rotation in the adsorbed layer
\cite{alkalionAg111,raregasonAg111,krongraphitePRL,krongraphitePRB,novacomctague,hexhex},
and a disruption of the normal periodicity and growth in the film
\cite{amorphousAu,starfishAl,fibonacciCu}. Domain wall systems in
monolayers can be thought of as some of the first examples of ordering
at the nanoscale, where the length scale is defined by the lattice
mismatch, and interestingly, mismatched systems are now designed and
used for growing nanostructure arrays having specific symmetries and
spacings.

After the discovery of quasicrystals, an interest grew in the
possibility of using quasicrystal surfaces as substrates to grow
quasicrystalline films of a single element, which do not occur in
nature \cite{mcgrathreview}.  Such a system necessarily has a lattice
mismatch between the adsorbate and the substrate, and the effects are
seen in the diverse examples of film structures and growth modes that
occur for these systems
\cite{amorphousAu,starfishAl,fibonacciCu,quantumAgBi,epitaxialBiSb,rotepitaxyAl,rotepitaxyAg}.
There is now interest in being able to design and produce
quasiperiodic arrays of nanoclusters on quasicrystal surfaces, but the
complex interactions involved in the adsorbate-substrate interactions
make it difficult to design specific nanostructures.

Recently, our group has begun to explore the behavior of simple gases
physisorbed on quasicrystalline surfaces \cite{XeQCPRL, XeQCPhilMag,
XeQCPRB, NobGasQCPhilMag}.  The key questions motivating these studies
are analogous to those for periodic substrates; for example, what are
the energy scales of the adsorbed film and what is the resulting
structure of the film?  The latter problem includes a specific
question: under what conditions, if any, of temperature $(T)$ and
vapor pressure $(P)$ does the film form an epitaxial phase on the
surface and when does it form an alternative structure, a hexagonal,
close-packed array of atoms or molecules (i.e., the film's ground
state within the two-dimensional (2D) approximation).  One expects the low $P$,
submonolayer behavior to be that of a film conforming to the
substrate, attracted to the sites offering the highest binding
energies.  The hypothetical behavior at high $P$ includes several
possibilities, e.g. bulk-like film growth, amorphous structures,
island formation or the absence of any multilayer film.  Which of
these scenarios (or others) actually occurs is a function of the
gas-surface and gas-gas interactions.

In this article, we report the results of Grand Canonical Monte Carlo
(GCMC) \cite{gcmc1,gcmc2} simulations of Ne, Ar, Kr, and Xe adsorbed
on the tenfold surface of a decagonal Al$_{73}$Ni$_{10}$Co$_{17}$
quasicrystal \cite{qc1,qc2}. We organize our article as follows. In
Section \ref{section_method} we discuss briefly the method. Section
\ref{section_results} is devoted to the results of Ne,
Ar, Kr, and Xe adsorbed on the AlNiCo. In Section
\ref{section_discussion} we compare the results of all gases. Section
\ref{section_conclusion} contains conclusions and comments on
strategies for future research in this area.

\section{Method}
\label{section_method}

By using the Grand Canonical Monte Carlo (GCMC) simulation method
\cite{gcmc3,gcmc1,gcmc2} we study the adsorption of noble gases: Ne,
Ar, Kr, and Xe on the tenfold surface of a decagonal
Al$_{73}$Ni$_{10}$Co$_{17}$ quasicrystal \cite{qc1,qc2} (He is omitted
because it requires a quantum treatment and Rn is omitted because it is not convenient experimentally). In this article, we use the abbreviation QC
to refer to this particular quasicrystalline substrate. GCMC, a widely
used method, is described in detail in references
\cite{XeQCPhilMag,XeQCPRB, NobGasQCPhilMag,scgcmc2}. Only a brief
overview is given here.

\subsection{Grand Canonical Monte Carlo}

At constant temperature, $T$, and volume, $V$, the GCMC method
explores the configurational phase space using the Metropolis
algorithm and finds the equilibrium number of adsorbed atoms
(adatoms), $N$, as a function of the chemical potential, $\mu$, of the
gas.  The adsorbed atoms are in equilibrium with the coexisting gas:
the chemical potential of the gas is constant throughout the system.
In addition, the coexisting gas is taken to be ideal.  With this
method we determine adsorption isotherms, $\rho_N$, and density
profiles, $\rho(x,y)$, as a function of the pressure, $P(T,\mu)$.  For
each data point in an isotherm, we perform at least 18 million GCMC
steps to reach equilibrium. Each step is an attempted displacement,
creation, or deletion of an atom with execution probabilities equal to
0.2, 0.4, and 0.4, respectively \cite{XeQCPRB,
NobGasQCPhilMag,scgcmc2}. At least 27 million steps are performed in the
subsequent data-gathering and -averaging phase.

\subsection{Unit cell}

The unit cell is tetragonal. We take a square section of the surface,
$A$, of side 5.12 nm, to be the $(x,y)$ part of the unit cell in the
simulation, for which we assume periodic boundary conditions along the
basal directions.  Although this assumption limits the accuracy of the
long range QC structure, it is numerically necessary for these
simulations. Since the size of the cell is relatively large compared
to that of the noble gases, the cell is accurately representative of
order on short-to-moderate length scales.  A hard wall at 10 nm above
the surface along $z$ is used to confine the coexisting vapor phase.
The simulation results for Xe over QC, presented below, are consistent
with both our results from experiments \cite{ref19} and virial
calculations \cite{ref18}. Hence, the calculations may also be accurate
for other systems.

\subsection{Gas-gas and gas-surface interactions}

The gas-gas potentials are taken to be Lennard-Jones (LJ) 12-6
interactions, with the parameter values $\epsilon_{gg}$ and
$\sigma_{gg}$ listed in table \ref{table_LJ}.  The gas-surface
potentials are based on a summation of two-body interactions between
the gas and the individual constituent atoms of the substrate: Al, Ni
and Co \cite{XeQCPRL,qc1,ref18}.  The gas-surface pair interactions
are also assumed to have LJ form, with parameter values taken from
traditional combining rules, using atomic sizes derived from bulk
crystalline lattice constants \cite{ref19,ref18,ref21,refZepp}. The LJ
gas-surface parameters are $\epsilon_{gas-Al}$ and $\sigma_{gas-Al}$
for Al, and $\epsilon_{gas-TM}$ and $\sigma_{gas-TM}$ for the two
transition metals Ni and Co. All these values are listed in the upper
part of table \ref{table_LJ}.  In the calculation of the adsorption
potential, we assume a structure of the unrelaxed surface taken from
the empirical fit to {\small LEED} data \cite{qc2}.

\begin{table}
\caption{
\label{table_LJ}
Parameter values for the 12-6 Lennard-Jones interactions. TM is the
label for Ni or Co. The prefixes i and d refer to hypothetical
inflated and deflated variants of real atoms, as discussed in the
text.
}
{\small
\begin{indented}
\item[]\begin{tabular}{ccccccc}
  \br
  \hline \hline
   & $\epsilon_{gg}$ & $\sigma_{gg}$ & $\epsilon_{gas-Al}$ & $\sigma_{gas-Al}$ & $\epsilon_{gas-TM}$ & $\sigma_{gas-TM}$ \\
   & (meV)           & (nm)          & (meV)               & (nm)              & (meV)               & (nm)              \\
      \hline
      Ne & 2.92 & 0.278 & 9.40 & 0.264 & 9.01 & 0.249 \\
      Ar & 10.32 & 0.340 & 17.67 & 0.295 & 16.93 & 0.280 \\
      Kr & 14.73 & 0.360 & 21.11 & 0.305 & 20.23 & 0.290 \\
      Xe & 19.04 & 0.410 & 24.00 & 0.330 & 23.00 & 0.315 \\
      \hline
      iNe$^{(1)}$ & 2.92 & 0.410 & 5.45 & 0.330 & 5.22 & 0.315 \\
      dXe$^{(1)}$ & 19.04 & 0.278 & 41.39 & 0.264 & 39.67 & 0.249 \\
      \hline
      dXe$^{(2)}$ & 19.04 & 0.390 & 25.88 & 0.320 & 24.80 & 0.305 \\
      iXe$^{(1)}$ & 19.04 & 0.550 & 14.96 & 0.400 & 14.34 & 0.385 \\
      iXe$^{(2)}$ & 19.04 & 0.675 & 10.52 & 0.462 & 10.08 & 0.447 \\
      \hline \hline 
      \end{tabular}
\end{indented}}
\end{table}

\subsection{Adsorption potentials}

\begin{figure}
 \centerline{
\epsfig{file=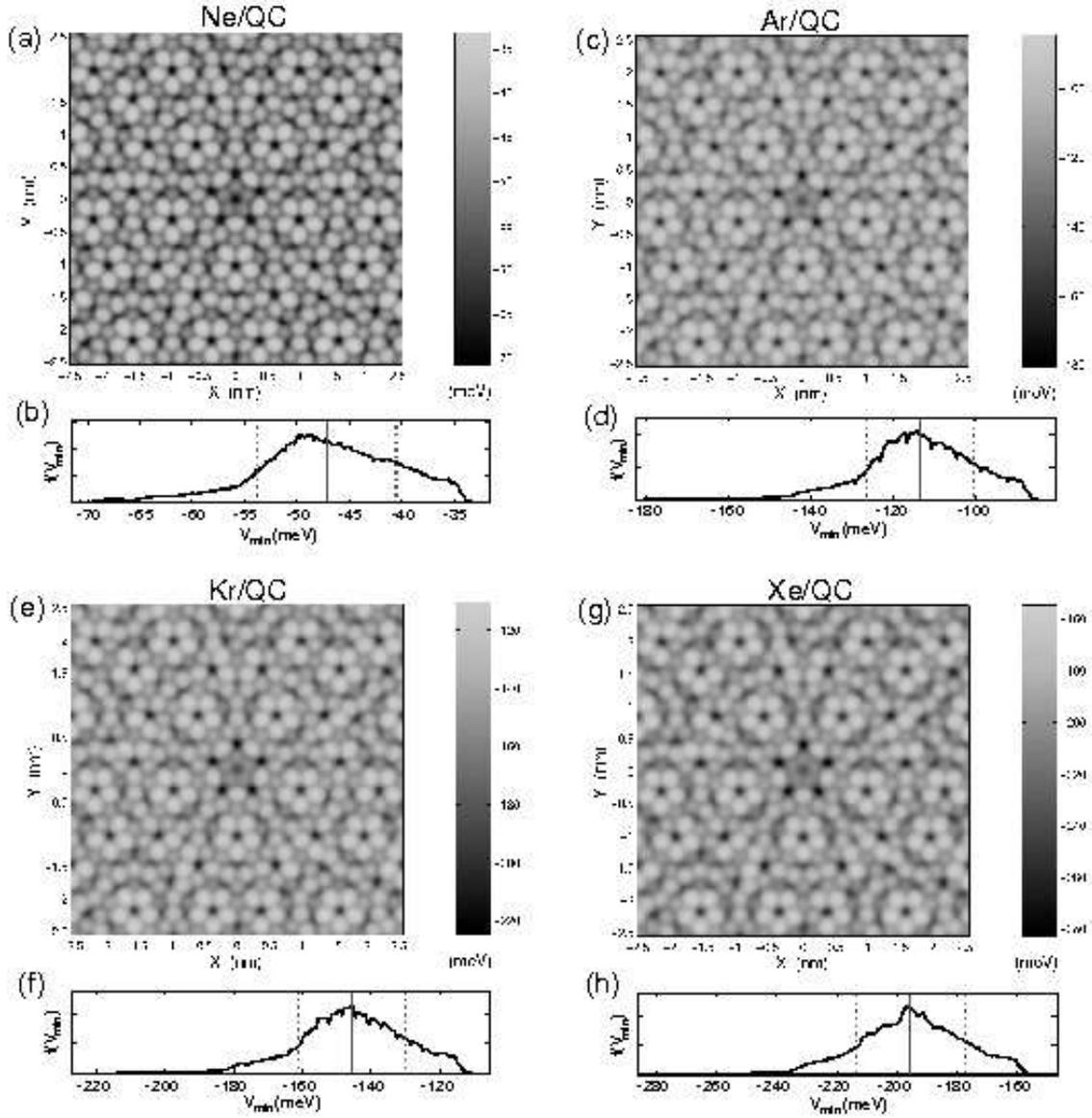,height=155mm,clip=}
}
 \vspace{1mm}
 \caption{\small
   (color online). Computed adsorption potentials for
   (a) Ne, (c) Ar, (e) Kr, and (g) Xe on the quasicrystal, 
   obtained by minimizing $V(x,y,z)$ with respect to $z$. 
   The distribution of the minimum value of these potentials is
   plotted in ((b), (d), (f), and (h)) respectively: the solid line marks the
   average value $\langle V_{min}\rangle $, the dashed lines mark the values at
   $\langle V_{min}\rangle \pm$SD.}
 \label{fig_potMINandHist}
\end{figure}

Figures \ref{fig_potMINandHist}(a), (c), (e), and (g) show the function
$V_{min}(x,y)$ of Ne, Ar, Kr, and Xe on the QC, respectively, which
is calculated by minimizing the adsorption potentials, $V(x,y,z)$,
along the $z$ direction at every value of the coordinates $(x,y)$:

\begin{equation} 
        \left. V_{min}(x,y) \equiv
        min\left\{V(x,y,z)\right\}\right|_{along\,\,z}.
\end{equation}
The figures reveal the fivefold rotational symmetry of the substrate.
Dark spots correspond to the most attractive regions of the substrate.
By choosing appropriate sets of five dark spots, we can identify
pentagons, whose sizes follow the inflationary property of the QC.
Note the pentagon at the center of each figure: it will be used to
extract the geometrical parameters $\lambda_s$ and $\lambda_c$ used in
Section \ref{section_discussion}.

To characterize the corrugation, not well-defined for aperiodic
surfaces, we calculate the distribution function $f(V_{min})$, the
average $\langle V_{min}\rangle $ and standard deviation SD of $V_{min}(x,y)$
as:
\begin{eqnarray}
  && f(V_{min}) dV_{min} \equiv probability \bigg\{ V_{min} \in \left[V_{min}, V_{min}+dV_{min} \right[ \bigg\}     \label{eq_fvminNobQC}\\
  && \langle V_{min}\rangle \equiv\int_{-\infty}^\infty f(V_{min})V_{min}\,dV_{min},      \label{eq_vminaveNobQC}\\
  && SD^2\equiv\int_{-\infty}^\infty f(V_{min})(V_{min}-\langle V_{min}\rangle )^2\,dV_{min}.      \label{eq_SDvminNobQC}
\end{eqnarray}
Figures \ref{fig_potMINandHist}(b), (d), (f), and (h) show $f(V_{min})$
of the adsorption potential for Ne, Ar, Kr, and Xe on the QC, respectively. $V_{min}(x,y)$
extends by more than $2\cdot$SD around its average, revealing the high
corrugation of the gas-surface interaction in these four systems. The
average and SD of $V_{min}(x,y)$ for these systems are listed in the
upper part of table \ref{table_NobGasQCPot}.

\subsection{Effective parameters}

For every gas-surface interaction we define two effective parameters
$\sigma_{gs}$ and $D_{gs}$.  $\sigma_{gs}$ represents the averaged LJ
size parameter of the interaction, calculated following the traditional
combining rules \cite{ref21}:

\begin{equation}
   \sigma_{gs}\equiv x_{Al}\sigma_{g-Al} + x_{Ni}\sigma_{g-Ni} +
   x_{Co}\sigma_{g-Co},
   \label{sigmags}
\end{equation}
where $x_{Al}$, $x_{Ni}$, and $x_{Co}$ are the concentrations of Al, Ni,
and Co in the QC, respectively.  $D_{gs}$ represents the
well depth of the laterally averaged potential $V(z)$:

\begin{equation}
 D_{gs} \equiv \left. -min\left\{V(z)\right\}\right|_{along\,\,z}.
   \label{epsilongs}
\end{equation}
In addition, we normalize the $\sigma_{gs}$ and $D_{gs}$ with
respect to the gas-gas interactions:
\begin{eqnarray}
  \sigma^\star_{gs}&\equiv&\sigma_{gs}/\sigma_{gg}, \\
  D^\star_{gs}&\equiv&D_{gs}/\epsilon_{gg}.
\end{eqnarray}
The values of the effective parameters $\sigma_{gs}$, $D_{gs}$,
$\sigma^\star_{gs}$, and $D^\star_{gs}$ for the four gas-surface
interactions are listed in the upper part of table \ref{table_NobGasQCPot}.  We
also include the well depth for Ne, Ar, Kr, and Xe on graphite, as
comparison \cite{vidali}.

\begin{table}
  \caption{\label{table_NobGasQCPot}Range, average ($\langle V_{min}\rangle $), and standard deviation (SD) of the interaction $V_{min}(x.y)$ on the QC.
    Effective parameters of the gas-surface interactions ($D_{gs}$, $\sigma_{gs}$, $D^\star_{gs}$,
    $\sigma^\star_{gs}$), and, for comparison, the best estimated well depths $D^{Gr}_{gs}$ on graphite \cite{vidali}.}
\begin{indented}
\item[]
{\scriptsize
        \begin{tabular}{ccccccccc}
        \br
        \hline \hline 
	& $V_{min}$ range & $\langle V_{min} \rangle$ & SD & $D_{gs}$ & $\sigma_{gs}$ & $D^\star_{gs}$             & $\sigma^\star_{gs}$ & $D^{Gr}_{gs}$ \\
	& (meV) & (meV) & (meV)       & (meV)    & (nm)          & $(D_{gs}/\epsilon_{gg})$ & $(\sigma_{gs}/\sigma_{gg})$ & (meV) \\
	\hline 
	Ne & -71 to -33 & -47.43 & 6.63 & 43.89 & 0.260 & 15.03 & 0.935 & 33\\
	Ar & -181 to -85 & -113.32 & 13.06 & 108.37 & 0.291 & 10.50 & 0.856 & 96\\ 
	Kr & -225 to -111 & -145.71 &  15.68 & 140.18 & 0.301 & 9.52 & 0.836 & 125\\
	Xe & -283 to -155 & -195.46 &  17.93 & 193.25 & 0.326 & 10.15 & 0.795 & 162\\
	\hline 
	iNe$^{(1)}$ & -65 to -36 & -45.11 & 4.08 & 43.89 & 0.326 & 15.03 & 0.795\\
	dXe$^{(1)}$ & -305 to -150 & -207.55 & 29.18 & 193.25 & 0.260 & 10.15 & 0.935\\
	\hline 
	dXe$^{(2)}$ & -295 to -155 & -199.40 & 19.33 & 193.25 & 0.316 & 10.15 & 0.810\\
	iXe$^{(1)}$ & -248 to -170 & -195.31 & 11.21 & 193.25 & 0.396 & 10.15 & 0.720\\
	iXe$^{(2)}$ & -230 to -180 & -194.25 & 7.77  & 193.25 & 0.458 & 10.15 & 0.679\\
	\hline
	\hline
    \end{tabular}
}
\end{indented}
\end{table}

\subsection{Fictitious gases}

As shown in tables \ref{table_LJ} and \ref{table_NobGasQCPot}, Ne is the {\it smallest}
atom and has the {\it weakest} gas-gas and gas-surface interactions (minima
of $\sigma_{gg}$, $\sigma_{gs}$, $\epsilon_{gg}$ and $D_{gs}$).  In
addition, Xe is the {\it largest} atom and has the {\it strongest} gas-gas and
gas-surface interactions (maxima of $\sigma_{gg}$, $\sigma_{gs}$,
$\epsilon_{gg}$ and $D_{gs}$).  Therefore, for our analysis, it is
useful to consider two ``fictitious'' gases, iNe$^{(1)}$ and
dXe$^{(1)}$, which are combinations of Ne and Xe parameters. \\
iNe$^{(1)}$ represents an ``inflated'' version of Ne, having the
same gas-gas and average gas-surface interactions of Ne but the
geometrical dimensions of Xe:
\begin{eqnarray}
  &&\{\epsilon_{gg},D_{gs},D^\star_{gs}\}[{\rm iNe}^{(1)}] \equiv
  \{\epsilon_{gg},D_{gs},D^\star_{gs}\}[{\rm Ne}], \label{iNe1a}\\
  &&\{\sigma_{gg},\sigma_{gs},\sigma^\star_{gs}\}[{\rm iNe}^{(1)}]
  \equiv \{\sigma_{gg},\sigma_{gs},\sigma^\star_{gs}\}[{\rm
  Xe}].\label{iNe1b}
\end{eqnarray}
dXe$^{(1)}$ represents a ``deflated'' version of Xe, having the
same gas-gas and average gas-surface interactions of Xe but the
geometrical dimensions of Ne:
\begin{eqnarray}
  &&\{\epsilon_{gg},D_{gs},D^\star_{gs}\}[{\rm dXe}^{(1)}] \equiv
  \{\epsilon_{gg},D_{gs},D^\star_{gs}\}[{\rm Xe}], \label{dXe1a}\\
  &&\{\sigma_{gg},\sigma_{gs},\sigma^\star_{gs}\}[{\rm dXe}^{(1)}]
  \equiv \{\sigma_{gg},\sigma_{gs},\sigma^\star_{gs}\}[{\rm
  Ne}].\label{dXe1b}
\end{eqnarray}
The resulting LJ parameters for iNe$^{(1)}$ and dXe$^{(1)}$ are
summarized in the central parts of tables \ref{table_LJ} and \ref{table_NobGasQCPot}.
Furthermore, we also define three other fictitious versions
of Xe: dXe$^{(2)}$, iXe$^{(1)}$, and iXe$^{(2)}$ which have the same
gas-gas and average gas-surface interactions of Xe but deflated or
inflated geometrical parameters.  The last three fictitious gases will
be used in Section \ref{section_discussion}. The LJ parameters
for these gases are summarized in the lower parts of tables \ref{table_LJ} and \ref{table_NobGasQCPot}.
In simulating fictitious gases,
we implicitly rescale the substrate's strengths so that the resulting
adsorption potentials 
have the same $D_{gs}$ as the non-inflated or non-deflated ones (equations \ref{iNe1a} and \ref{dXe1a}).

\subsection{Chemical potential, order parameter, and ordering transition} 
\label{section_orderdefinition}

To appropriately characterize the evolution of the adsorption processes of
the gases we define a {\it reduced chemical potential} $\mu^\star$, as:
\begin{equation}
  \mu^\star \equiv \frac{\mu-\mu_1}{\mu_2-\mu_1},
  \label{eqmu}
\end{equation}
where $\mu_1$ and $\mu_2$ are the chemical potentials at the onset of
the first and second layer formation, respectively. 
In addition, as done in reference \cite{XeQCPRB, NobGasQCPhilMag},
we introduce the {\it order parameter} $\rho_{5-6}$, 
defined as the probability of existence of fivefold defect:
\begin{equation}
  \rho_{5-6} \equiv \frac{N_5}{N_5 + N_6},
  \label{eqrho}
\end{equation}
where $N_5$ and $N_6$ are the numbers of atoms having 2D coordination
equal to 5 and 6, respectively. The 2D coordination is the number of 
neighboring atoms within a cutoff radius of $a_{NN}\cdot
1.366$ where $a_{NN}$ is the first nearest neighbor (NN) distance of the gas
in the solid phase and $1.366=\cos(\pi/6)+1/2$ is the average of the
first and the second NN distances in a triangular lattice.

In a fivefold ordering, most arrangements are hollow or filled
pentagons with atoms having mostly five neighbors. Hence, the
particular choice of $\rho_{5-6}$ is motivated by the fact that such
pentagons can become hexagons by gaining additional atoms with five or
six neighbors. {\bf Definition:} the five to sixfold ordering
transition is defined as a decrease of the order parameter to a small
or negligible final value. The phenomenon can be abrupt (first-order)
or continuous. Within this framework,  $\rho_{5-6}$ and
($1-\rho_{5-6}$) can be considered as the fractions of pentagonal and
triangular phases in the film, respectively.

\section{RESULTS}
\label{section_results}

\subsection{Adsorption isotherms}
\label{section_isotherms}

\begin{figure}[htb]
\begin{indented}
\item[]
\epsfig{file=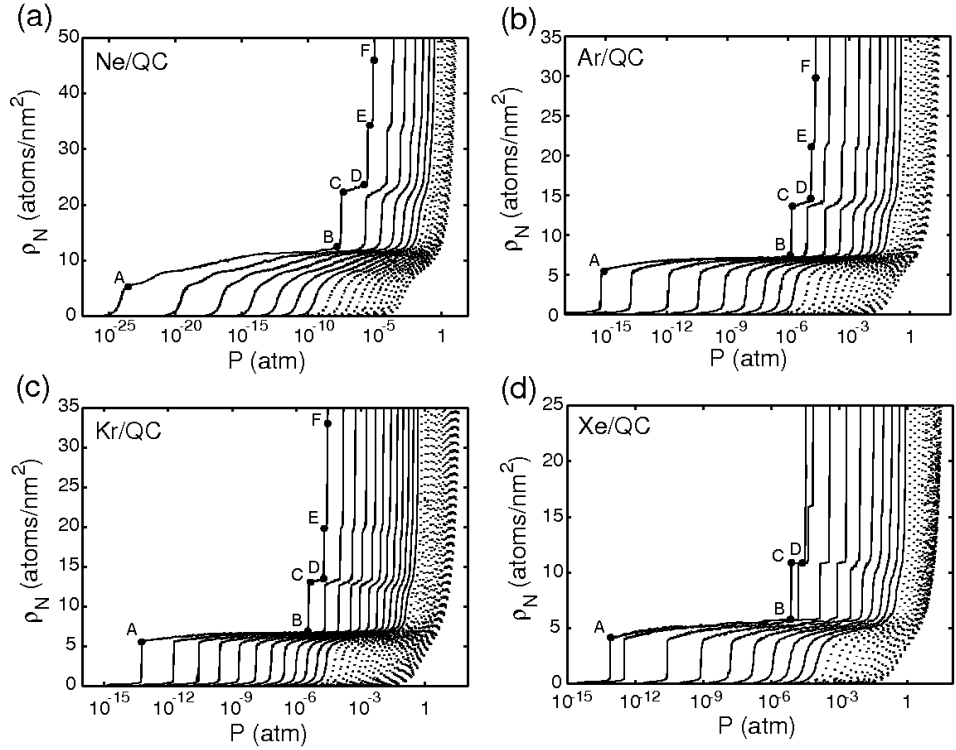,height=100mm,clip=}
 \vspace{1mm}
 \caption{\small
   Computed adsorption isotherms for all the gas/QC systems.  The
   ranges of temperatures under study are: 
   Ne: $T=$ 14 K to 46 K in 2 K steps, 
   Ar: 45 K to 155 K in 5 K steps, 
   Kr: 65 K to 225 K in 5 K steps, 
   Xe: 80 K to 280 K in 10 K steps.
   Additional isotherms are shown with solid circles at
   $T^\star=0.35$: $T=11.8$ K (Ne), $T=41.7$ K (Ar), $T=59.6$ K (Kr),
   and $T=77$ K (Xe).  Isotherms above the triple point temperatures
   are shown as dotted curves.}
 \label{fig_isotherms}
\end{indented}
\end{figure}

Figure \ref{fig_isotherms} shows the adsorption isotherms of Ne,
Ar, Kr, and Xe on the QC. The plotted quantities are the densities of
adatoms per unit area, $\rho_N$, as a function of pressure at various
temperatures. The simulated ranges and the experimental triple point
temperatures ($T_t$) for Ne, Ar, Kr, and Xe are listed in table
\ref{table_results}.  A layer-by-layer film growth is visible
at low temperatures. A complete wetting behavior is observed as
indicated by a continuous film growth at temperatures above $T_t$
(isotherms at $T>T_t$ are shown as dotted curves). This behavior, 
observed in spite of the high corrugation, is interesting
as corrugation has been shown to be capable of preventing wetting
\cite{refWahyu1,refWahyu2}.

Although vertical steps corresponding to layers' formation are evident
in the isotherms, the slopes of the isotherms' plateaus at the same
normalized temperatures ($T^\star\equiv T/\epsilon_{gg}=0.35$) differ
between systems.  To characterize this, we calculate the increase of
each layer density, $\Delta\rho_N$, from the formation to the onset of
the subsequent layer. $\Delta\rho_N$ is defined as $\Delta\rho_N
\equiv (\rho_B-\rho_A)/\rho_A$ and the values are reported in table
\ref{table_results} (points (A) and (B) are specified in figure
\ref{fig_isotherms}).  We observe that, as the size of noble gas
increases $\Delta\rho_N$ become smaller, indicating that the substrate
corrugation has a more pronounced effect on smaller adsorbates, as
expected since they penetrate deeper into the corrugation pockets.
However, Xe/QC does not follow this trend. This arises from the
complex interplay between the corrugation energy and length of the
potential with respect to the parameters of the gas
$(\sigma_{gg},\epsilon_{gg})$ in determining the density of the
adsorbed layers. In the case of Ne, Ar, and Kr on QC, the densities at
points (A) are approximately the same ($\rho_A$ = 5.4 atoms/nm$^2$),
whereas that of Xe/QC is considerably smaller ($\rho_A$ = 4.2
atoms/nm$^2$), because the Xe dimension $\sigma_{gg}$ becomes
comparable to the characteristic length (corrugation) of the
potential.  This effect is clarified by the density profile of the
films, $\rho(x,y)$, shown in figures \ref{fig_rhoNeAr} and
\ref{fig_rhoKrXe}.  As can be seen at points (A), the density profiles
of Ne, Ar, and Kr on QC are the same, i.e.  the same set of dark spots
appear in their plots.  For Xe, some spots are separated with
distances smaller than its core radius ($\sigma_{gg}$), causing
repulsive interactions. Hence these spots will not likely appear in
the density profile, resulting in a lower $\rho_A$.  More discussion
on how interaction parameters affect the shape of the isotherms is
presented in Section \ref{section_discussion}.  Note that the second
layer in each system has a smaller $\Delta\rho_N$ than the first
one. The explanation will be given when we discuss the evolution of
density profiles.

\begin{table}[htb]
  \caption{\label{table_results} Results for Ne, Ar, Kr, and
  Xe adsorbed on the QC. $T_t$ is taken from reference\cite{crawford}. The
  density increase ($\Delta\rho_N$) in the first and second layers is
  calculated at $T^\star=0.35$ from point (A) to (B) and (C) to (D)
  in figure \ref{fig_isotherms}, respectively.}
\begin{indented}
\item[]
{\scriptsize
\begin{tabular}{cccclc}
  \br
  \hline \hline
   & simulated $T$ & $T^\star\equiv T/\epsilon_{gg}$ & $T_t$ & \hspace{16 mm} $\Delta\rho_N$ at $T^\star=0.35$ & $\theta_r$\\
   & (K)           &                & (K)   & \hspace{5 mm} for 1$^{st}$ layer \hspace{13 mm} for 2$^{nd}$ layer   & \\
      \hline
      Ne & 11.8 $\rightarrow$ 46  & 0.35 $\rightarrow$ 1.36 & 24.55  & (12.2-5.3)/5.3=1.30 \hspace{2 mm} (11.1-10.2)/10.2=0.09 & 6$^\circ$\\
      Ar & 41.7 $\rightarrow$ 155 & 0.35 $\rightarrow$ 1.29 & 83.81  & (7.3-5.5)/5.5=0.33 \hspace{8 mm} (6.9-6.4)/6.4=0.08 & 30$^\circ$\\
      Kr & 59.6 $\rightarrow$ 225 & 0.35 $\rightarrow$ 1.32 & 115.76 & (6.9-5.5)/5.5=0.25 \hspace{8 mm} (6.6-6.3)/6.3=0.05 & 42$^\circ$\\
      Xe & 77 $\rightarrow$ 280   & 0.35 $\rightarrow$ 1.27 & 161.39 & (5.8-4.2)/4.2=0.38 \hspace{8 mm} (5.2-5.2)/5.2=0 & 54$^\circ$\\
      \hline \hline 
      \end{tabular}
}
\end{indented}
\end{table}

\subsection{Density profiles}
\label{density_profiles}

\begin{figure}
\begin{indented}
\item[]
\epsfig{file=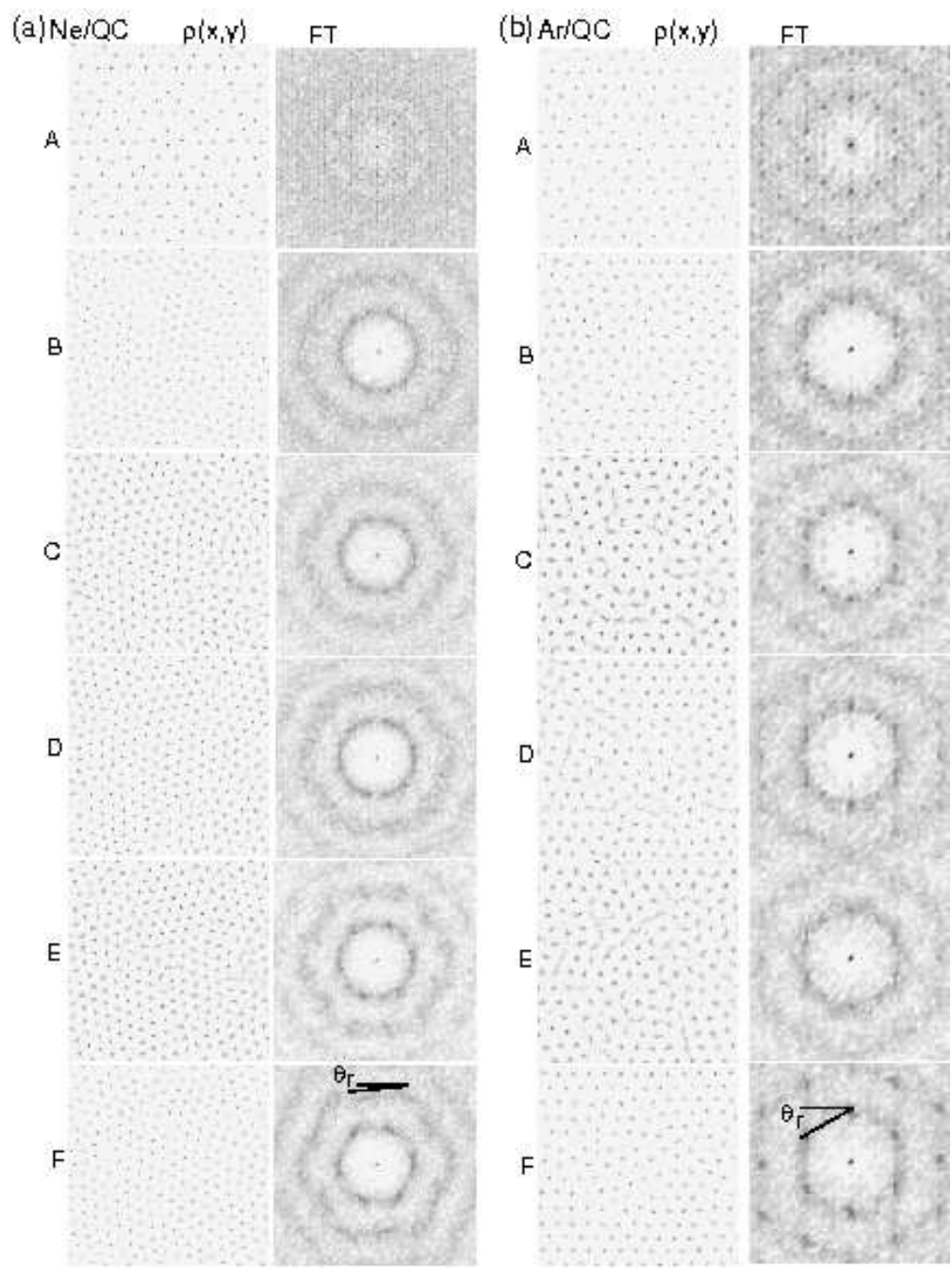,height=178mm,clip=}
 \vspace{1mm}
 \caption{\small
   Density profiles and Fourier transforms of the outer layer 
   at $T^\star=0.35$ for Ne/QC ($T=11.8$ K) and Ar/QC ($T=41.7$ K), 
   corresponding to points (A) through (F) of figure \ref{fig_isotherms}.}
 \label{fig_rhoNeAr}
\end{indented}
\end{figure}

\begin{figure}
\begin{indented}
\item[]
\epsfig{file=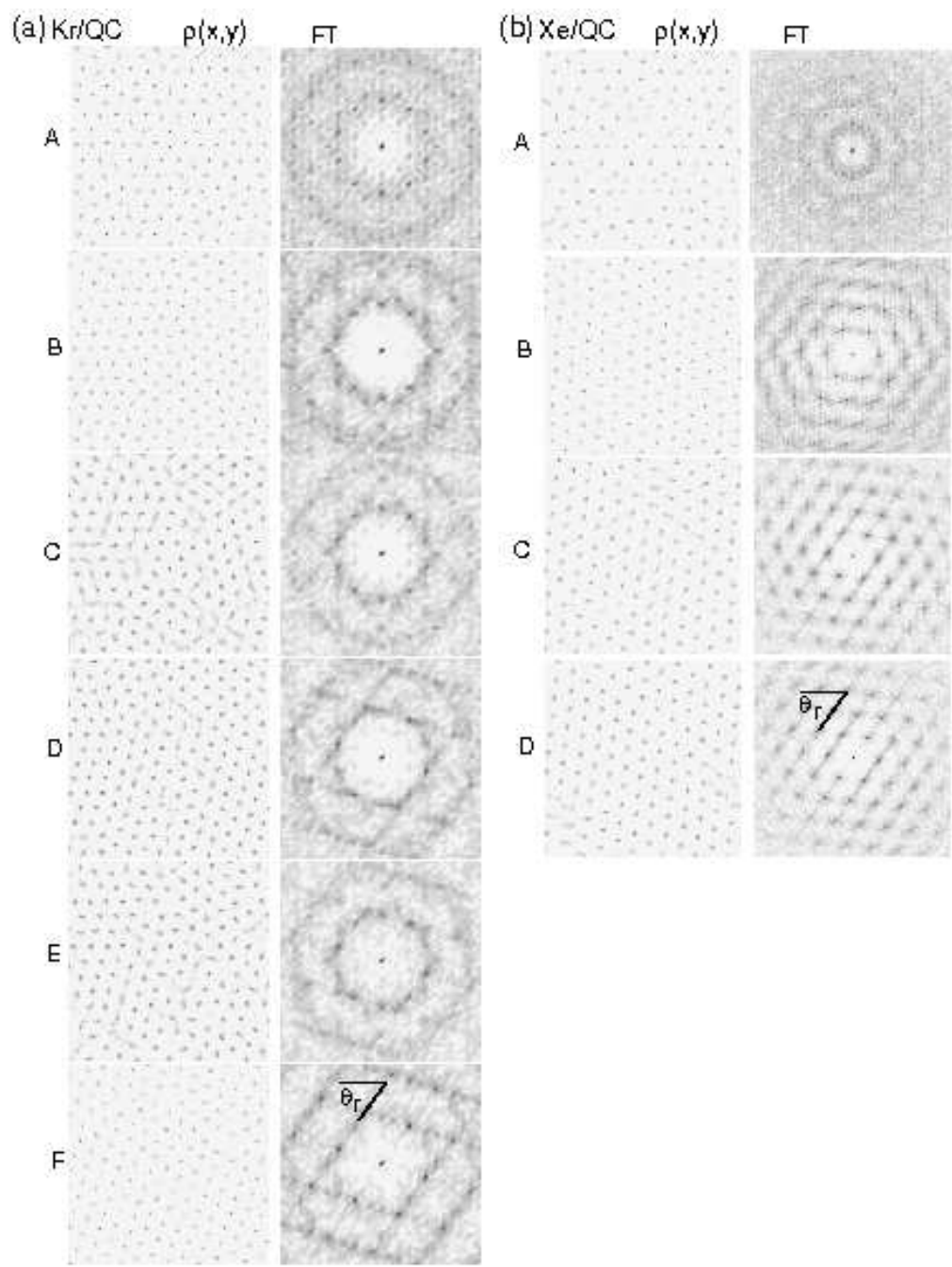,height=178mm,clip=}
 \vspace{1mm}
 \caption{\small
   Density profiles and Fourier transforms of the outer layer 
   at $T^\star=0.35$ for of Kr/QC ($T=59.6$ K) and Xe/QC ($T=77$ K),
   corresponding to points (A) through (F) of figure \ref{fig_isotherms}.}
 \label{fig_rhoKrXe}
\end{indented}
\end{figure}

Figures \ref{fig_rhoNeAr} and \ref{fig_rhoKrXe} show the density profiles
$\rho(x,y)$ at $T^\star=0.35$ for the outer layers of Ne, Ar,
Kr, and Xe adsorbed on the QC at the pressures corresponding to points (A) through
(F) of the isotherms in figure \ref{fig_isotherms}.

{\bf Ne/QC system.}  Figure \ref{fig_rhoNeAr}(a) shows the evolution of
adsorbed Ne.  At the formation of the first layer, adatoms are
arranged in a pentagonal manner following the order of the substrate,
as shown by the discrete spots of the Fourier transform (FT) having
tenfold symmetry (point (A)).  As the pressure increases, the
arrangement gradually loses its pentagonal character.  In fact, at
point (B) the adatoms are arranged in patches of triangular lattices
and the FT consists of uniformly-spaced concentric rings with
hexagonal resemblance.  The absence of long-range ordering in the
density profile is indicated by the lack of discrete spots in the
FT. This behavior persists throughout the formation of the second
layer (points (C) and (D)) until the appearance of the third layer
(point (E)).  At this and higher pressures, the FT shows patterns
oriented as hexagons rotated by $\theta_r$ = 6$^\circ$, indicating the
presence of short-range triangular order on the outer layer (point
(F)).  In summary, between points (A) and (F) the arrangement evolves
from pentagonal fivefold to triangular sixfold with considerable
disorder, as the upper part of the density profile at point (F) shows.
The transformation of the density profile, from a
lower-packing-density (pentagonal) to a higher-packing-density
structure (irregular triangular), occurs mostly in the monolayer from
points (A) to (B), causing the largest density increase of the first
layer with respect to that of the other layers (see the end of Section
\ref{density_profiles} for more discussion). Due to the considerable
amount of disorder in the final state Ne/QC does not satisfy the
requirements for the transition as defined in Section
\ref{section_orderdefinition}.

{\bf Ar/QC and Kr/QC systems.}  Figures \ref{fig_rhoNeAr}(b) and
\ref{fig_rhoKrXe}(a) show the evolutions of Ar/QC and Kr/QC: they are
similar to the Ne/QC system.  For Ar/QC, the pentagonal structure
at the formation of the first layer is confirmed by the FT showing
discrete spots having tenfold symmetry (point (A)).  The quasicrystal
symmetry strongly affects the overlayers' structures up to the third
layer by preventing the adatoms from forming a triangular lattice
(point (E)).  This appears, finally, in the lower part of the density
profile at the formation of the fourth layer as confirmed by the FT
showing discrete spots with sixfold symmetry (point (F)).  Similarly
to the Ne/QC system, disorder does not disappear but remain
present in the middle of the density profile corresponding to the
highest coverage before saturation (point (F)).  Similar situation occurs also
for the evolution of Kr/QC as shown in figure \ref{fig_rhoKrXe}(a).

{\bf Xe/QC system.}  Figure \ref{fig_rhoKrXe}(b) shows the evolution of
adsorbed Xe.  At the formation of the first layer, adatoms are
arranged in a fivefold ordering similar to that of the substrate as
shown by the discrete spots of the FT having tenfold symmetry (point
(A)). At point (B), the density profile shows a well-defined
triangular lattice not present in the other three systems: the FT
shows discrete spots arranged in regular and equally-spaced concentric
hexagons with the smallest containing six clear spots.  Thus, at point
(B) and at higher pressures, the Xe overlayers can be considered to
have a regular closed-packed structure with negligible
irregularities.

It is interesting to compare the orientation of the hexagons on FT for
these four adsorbed gases at the highest available pressures
before saturation (point (F) for Ne, Ar, and Kr, and point (D)
for Xe).  We define the orientation angles as the smallest of the
possible clockwise rotations to be applied to the hexagons to obtain
one side horizontal, as shown in figures \ref{fig_rhoNeAr} and
\ref{fig_rhoKrXe}.  Such angles are $\theta_r$ = 6$^\circ$, 30$^\circ$,
42$^\circ$, and 54$^\circ$, for adsorbed Ne, Ar, Kr, and Xe,
respectively.  These orientations, induced by the fivefold symmetry of
the QC, can differ only by multiples of $n\cdot 12^\circ$
\cite{XeQCPRB, NobGasQCPhilMag}.  Since hexagons have sixfold
symmetry, our systems can access only five possible orientations
($6,18,30,42,54^\circ$), and the final angles are determined by the
interplay between the adsorbate solid phase lattice spacing, the
periodic simulation cell size, and the potential corrugation.  For
systems without periodic boundary conditions, the ground state has
been found to be fivefold degenerate, as should be the case
\cite{XeQCPRB, NobGasQCPhilMag}.

In every system, the increase of the density for each layer is
strongly correlated to the commensurability with its support: the more
similar they are, the more flat the adsorption isotherm will be (note
that the support for the $(N+1)^{th}$-layer is the $N^{th}$-layer).
For example, the Xe/QC system has an almost perfect hexagonal
structure at point (B) (due to its first-order five to sixfold
ordering transition as described in the next section).  Hence, all the
further overlayers growing on the top of the monolayer will be at
least {\it ``as regular''} as the first layer, and have the negligible
density increase as listed in table \ref{table_results}.

\subsection{Order parameters}
\label{section_orderparameter}

\begin{figure}
\begin{indented}
\item[]
\epsfig{file=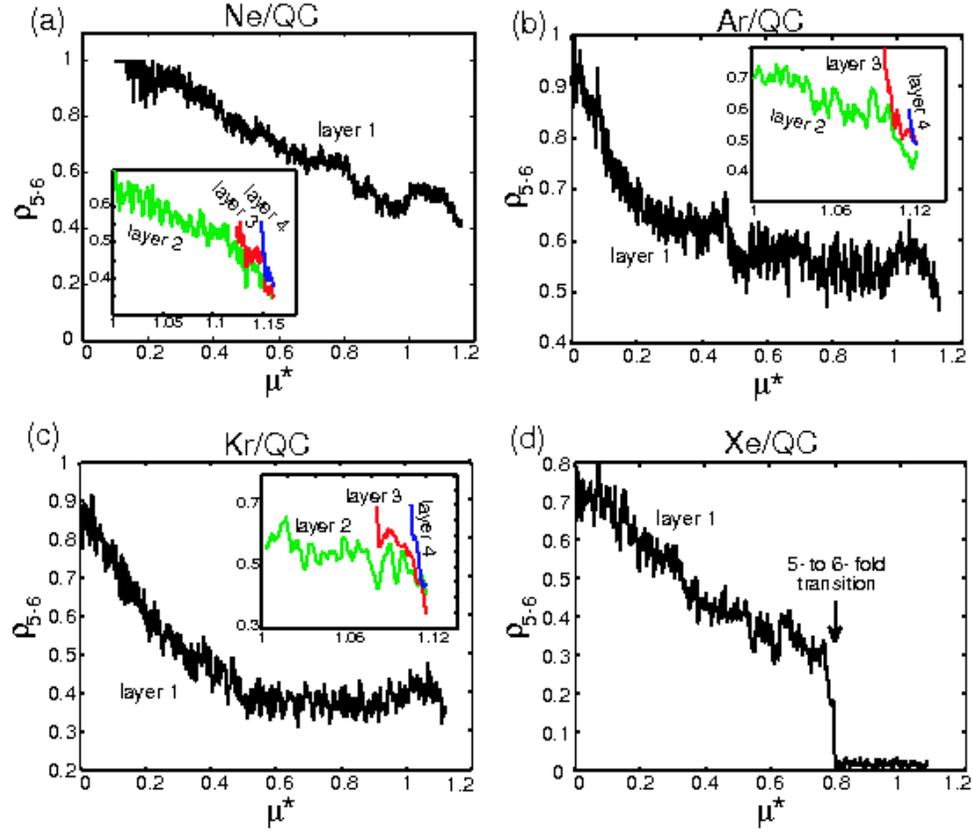,width=130mm,clip=}
 \vspace{1mm}
 \caption{\small
  (color online). Order parameters, $\rho_{5-6}$, as a function of
  normalized chemical potential, $\mu^\star$, (as defined in the text)
  at $T^\star=0.35$ for the first four layers of (a) Ne, (b) Ar,
  (c) Kr, and for the first layer of Xe (d) adsorbed on the quasicrystal. A sudden drop of
  the order parameter in Xe/QC to a constant value of $\sim0.017$ at
  $\mu^\star\sim 0.8$ indicates the existence of a first-order
  structural transition from fivefold to sixfold in the system.}
 \label{fig_df56}
\end{indented}
\end{figure}

The evolution of the order parameter $\rho_{5-6}$ is shown in
figure \ref{fig_df56} as a function of the normalized chemical 
potential, $\mu^\star$, at $T^\star$=0.35 for all the noble gases/QC systems.

{\bf Ne/QC, Ar/QC, and Kr/QC systems.} 
The $\rho_{5-6}$ plots for the first four layers observed before bulk
condensation are shown in panels (a)$-$(c).  As the chemical potential
$\mu^\star$ increases, $\rho_{5-6}$ decreases continuously reaching a
constant value only for Kr/QC.  At bulk condensation, the values of
$\rho_{5-6}$ are still high, approximately 0.35 $\sim$ 0.45. Data at
higher temperatures shows a similar behavior (up to $T$=24 K
($T^\star$=0.71) for Ne, $T=$70 K ($T^\star$=0.58) for Ar, and $T$=90
K ($T^\star$=0.53) for Kr). Thus, we conclude that these systems do
not undergo the ordering transition.

{\bf Xe/QC system.} The $\rho_{5-6}$ plot for the first layer is shown
in panel (d). In this system, as the chemical potential $\mu^\star$
increases, the order parameter gradually decreases reaching a value of
$\sim$ 0.3 at $\mu_{tr}^\star \sim$ 0.8. Suddenly it drops to 0.017
and remains constant until bulk condensation.  Similar behavior is
observed at higher temperatures up to $T$=140 K ($T^\star$=0.63).
This is a clear indication of a five to sixfold ordering transition,
as the first layer has undergone a transformation to an almost perfect
triangular lattice.  In addition, the transition has been found to
have temperature-dependent critical chemical potential
($\mu_{tr}(T)$), and to be first order with associated latent
heat\cite{XeQCPRB, NobGasQCPhilMag}.

\section{Discussion}
\label{section_discussion}

\begin{figure}
\begin{indented}
\item[]
\epsfig{file=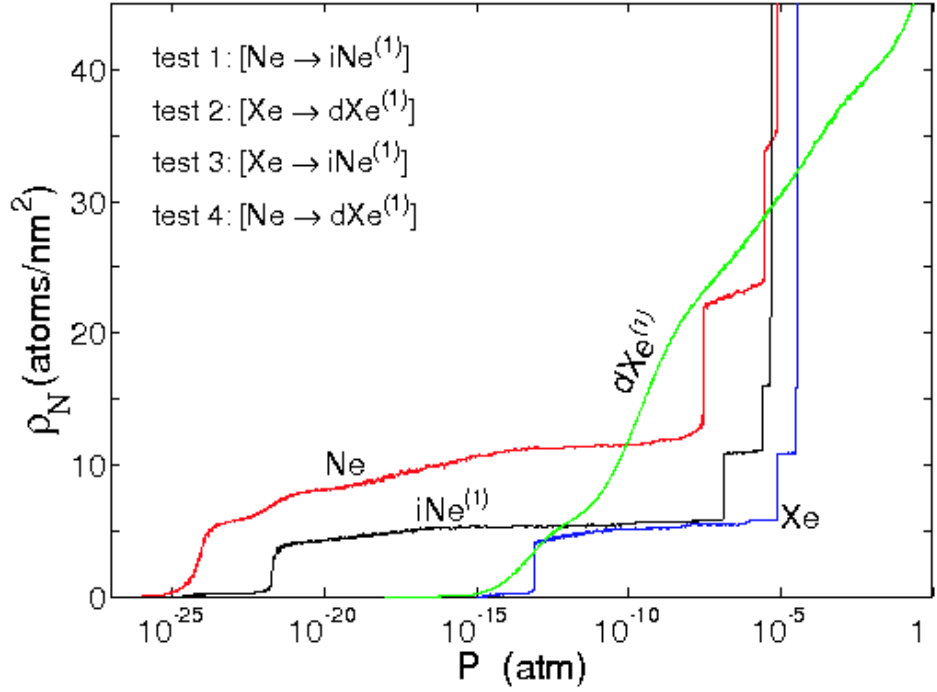,height=95mm,clip=}
 \caption{\small
  (color online). Computed adsorption isotherms for Ne, Xe, iNe$^{(1)}$, and
  dXe$^{(1)}$ on the quasicrystal at $T^\star$=0.35. iNe$^{(1)}$ and dXe$^{(1)}$ are
  fictitious noble gases having potential parameters described in the
  text and in tables \ref{table_LJ} and \ref{table_NobGasQCPot}.
  The effect of varying the interaction strength of the adsorbates on the
  density increase $\Delta\rho_N$ (while keeping the size constant) 
  is negligible on large gases but significant on small gases.}
 \label{fig_isothermNeXeG1G2}
\end{indented}
\end{figure}

In Section \ref{section_isotherms} we have briefly discussed how the
density increase of each layer ($\Delta\rho_N$) is affected by the
size of the adsorbate ($\sigma_{gg}$).  In addition, since the
corrugation of the potential depends also on the gas-gas interaction
($\epsilon_{gg}$), the latter quantity could {\it a priori} have an
effect on the density increase.  To decouple the effects
of $\sigma_{gg}$ and $\epsilon_{gg}$ on $\Delta\rho_N$ we calculate
$\Delta\rho_N$ while keeping one parameter constant, $\sigma_{gg}$ or
$\epsilon_{gg}$, and varying the other.  For this purpose, we
introduce two fictitious gases iNe$^{(1)}$ and dXe$^{(1)}$, which
represent ``inflated'' or ``deflated'' versions of Ne and Xe,
respectively (parameters are defined in equations
\ref{iNe1a}-\ref{dXe1b} and listed in tables \ref{table_LJ} and
\ref{table_NobGasQCPot}).  Then we perform four tests summarized as the following:
\def\testONE{[Ne$\rightarrow$iNe$^{(1)}$]}
\def\testTWO{[Xe$\rightarrow$dXe$^{(1)}$]}
\def\testTHREE{[Xe$\rightarrow$iNe$^{(1)}$]}
\def\testFOUR{[Ne$\rightarrow$dXe$^{(1)}$]}
\begin{itemize}
\item[]{\bf (1)} constant strength $\epsilon_{gg}$, size $\sigma_{gg}$ increases \testONE: $\Delta\rho_N$ reduces, 
\item[]{\bf (2)} constant strength $\epsilon_{gg}$, size $\sigma_{gg}$ decreases \testTWO: $\Delta\rho_N$ increases, \item[]{\bf (3)} constant size $\sigma_{gg}$, strength $\epsilon_{gg}$ decreases \testTHREE: $\Delta\rho_N$ $\sim$ constant,
\item[]{\bf (4)} constant size $\sigma_{gg}$, strength $\epsilon_{gg}$ increases \testFOUR: enhanced agglomeration.
\end{itemize}
Figure \ref{fig_isothermNeXeG1G2} shows the adsorption isotherms at
$T^\star$=0.35 for Ne, iNe$^{(1)}$, Xe, and dXe$^{(1)}$ on the QC.
By keeping the strength constant and varying the size of the adsorbates,
tests 1 and 2 (\testONE\, and \testTWO), we find that we can reduce or
increase the value of the density increase (when $\Delta\rho_N$
decreases the continuous growth tends to become stepwise and vice
versa).  These two tests indicate that the larger the size, the
smaller the $\Delta\rho_N$. By keeping the size constant and
decreasing the strength, test 3 (\testTHREE), we find that
$\Delta\rho_N$ does not change appreciably.  An interesting phenomenon
occurs in test 4 where we keep the size constant and increase the
strength (\testFOUR).  In this test the film's growth loses its
step-like shape.  We suspect that this is caused by an enhanced
agglomeration effect as follows.  Ne and dXe$^{(1)}$ have the same
size which is the smallest of the simulated gases,
allowing them to easily follow the substrate corrugation, in which
case, the corrugation helps to bring adatoms closer to each other
\cite{ref18} (agglomeration effect). The stronger gas-gas self
interaction of dXe$^{(1)}$ compared to Ne will further enhance this
agglomeration effect, resulting in a less stepwise film growth of
dXe$^{(1)}$ than Ne. As can be seen, dXe$^{(1)}$ grows continuously,
suggesting a strong enhancement of the agglomeration.  In summary, the
last two tests (3 and 4) indicate that the effect of varying the
interaction strength of the adsorbates (while keeping the size constant)
is negligible on large gases but significant on small gases.

Strength $\epsilon_{gg}$ and size $\sigma_{gg}$ of the adsorbates also
affect the existence of the first-order transition (present in
Xe/QC, but absent in Ne/QC, Ar/QC, and Kr/QC).  Hence we perform the
same four tests described before and observe the evolution of the
order parameter.  The results are the following:
\begin{itemize}
\item[]{\bf (1)} constant strength $\epsilon_{gg}$, size $\sigma_{gg}$ increases \testONE: transition appears,
\item[]{\bf (2)} constant strength $\epsilon_{gg}$, size $\sigma_{gg}$ decreases \testTWO: transition disappears,
\item[]{\bf (3)} constant size $\sigma_{gg}$, strength $\epsilon_{gg}$ decreases \testTHREE: transition remains,
\item[]{\bf (4)} constant size $\sigma_{gg}$, strength $\epsilon_{gg}$ increases \testFOUR: no transition appears.
\end{itemize}
The strength $\epsilon_{gg}$ has no effect on the existence of the
transition (tests 3 and 4), which instead is controlled by the size of
the adsorbates (tests 1 and 2).  To further characterize such dependence,
we add three additional fictitious gases with the same strength
$\epsilon_{gg}$ of Xe but different sizes $\sigma_{gg}$.  The three
gases are denoted as dXe$^{(2)}$, iXe$^{(1)}$, and iXe$^{(2)}$ (the
prefixes d- and i- stand for deflated and inflated, respectively).
The interaction parameters, defined in the following equations,
are listed in tables \ref{table_LJ} and \ref{table_NobGasQCPot}:
\begin{eqnarray}
  &&\left\{\epsilon_{gg},D_{gs},\sigma_{gg}\right\}[{\rm dXe}^{(2)}]
  \equiv \{\epsilon_{gg},D_{gs},0.95\sigma_{gg}\}[{\rm
  Xe}], \label{dXe2}\\
  &&\{\epsilon_{gg},D_{gs},\sigma_{gg}\}[{\rm iXe}^{(1)}]
  \equiv \{\epsilon_{gg},D_{gs},1.34\sigma_{gg}\}[{\rm
  Xe}], \label{iXe1}\\
  &&\{\epsilon_{gg},D_{gs},\sigma_{gg}\}[{\rm iXe}^{(2)}]
  \equiv \{\epsilon_{gg},D_{gs},1.65\sigma_{gg}\}[{\rm
  Xe}]. \label{iXe2}
\end{eqnarray}

\begin{figure}
\begin{indented}
\item[]
\epsfig{file=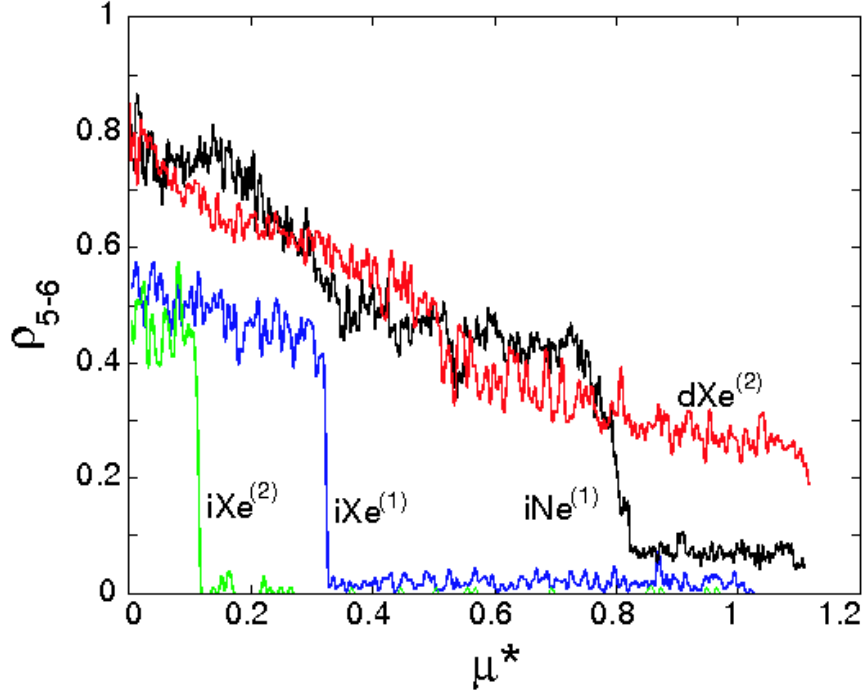,height=95mm,clip=}
 \caption{\small (color online). Order parameters as a function of
  normalized chemical potential (as defined in the text) for the first
  layer of dXe$^{(2)}$, iNe$^{(1)}$, iXe$^{(1)}$, and iXe$^{(2)}$ adsorbed on
  the quasicrystal at $T^\star=0.35$. A first-order fivefold to sixfold structural
  transition occurs in the last three systems, but not in
  dXe$^{(2)}$/QC.}
 \label{fig_df56iNe1iXe1dXe2}
\end{indented}
\end{figure}

Figure \ref{fig_df56iNe1iXe1dXe2} shows the evolutions of the order
parameter as a function of the normalized chemical potential for the
first layer of dXe$^{(2)}$, iNe$^{(1)}$, iXe$^{(1)}$, and iXe$^{(2)}$
adsorbed on the QC at $T^\star$=0.35. All these systems undergo a
transition, except dXe$^{(2)}$/QC, i.e. the transition occurs only in
systems with $\sigma_{gg}\geq\sigma_{gg}$[Xe] indicating the existence
of a critical value for the appearance of the phenomenon.
Furthermore, as $\sigma_{gg}$ increases (iNe$^{(1)}\rightarrow$
iXe$^{(1)}\rightarrow$ iXe$^{(2)}$), the transition shifts towards
smaller critical chemical potentials.

The critical value of $\sigma_{gg}$ can be related to the characteristic 
length of the QC by introducing a {\it gas-QC mismatch parameter} defined as
\begin{equation}
   \delta_m \equiv \frac{k\cdot\sigma_{gg}-\lambda_r}{\lambda_r}.
   \label{delta_m}
\end{equation}
where $k=0.944$ is the distance between rows in a close-packed plane
of a bulk LJ gas (calculated at $T=0$ K with $\sigma=1$
\cite{bruch2}), and $\lambda_r$ is the characteristic spacing of the
QC, determined from the momentum transfer analysis of LEED patterns
\cite{ref19} (our QC surface has $\lambda_r$=0.381 nm \cite{ref19}).
With such {\it ad hoc} definition, $\delta_m$ measures the mismatch
between an adsorbed \{111\} fcc plane of adatoms and the QC surface.  In
table \ref{table_mismatch} we show that $ \delta_m$ perfectly
correlates with the presence of the transition in our test cases
(transition exists $\Leftrightarrow$ $\delta_m>$ 0).

\begin{table}[htb]
\caption{
  \label{table_mismatch}
  Summary of adsorbed noble gases on the QC that undergo a
  first-order fivefold to sixfold structural transition and those
  that do not.}
  \begin{indented}
  \item[]\begin{tabular}{ccc}
  \br
    \hline \hline
       & $\delta_m$ & transition\\
     \hline
     Ne & -0.311 &  No\\
     Ar & -0.158 &  No\\
     Kr & -0.108 &  No\\
     Xe &  0.016 &  Yes\\
     iNe$^{(1)}$ &  0.016 &  Yes\\
     dXe$^{(1)}$ & -0.311 &  No\\
     dXe$^{(2)}$ & -0.034 &  No\\
     iXe$^{(1)}$ & 0.363 &  Yes\\
     iXe$^{(2)}$ & 0.672 &  Yes\\
     \hline \hline
   \end{tabular}
   \begin{tabular}{c}
     $k$ = 0.944 \cite{bruch2}\\
     $\lambda_r$ = 0.381 nm \cite{ref19}\\
     $\delta_m \equiv (k\cdot\sigma_{gg}-\lambda_r)/\lambda_r$\\
   \end{tabular}
   \end{indented}
\end{table}

The definition of a gas-QC mismatch parameter is not unique.  For
example, one can substitute $k\cdot\sigma_{gg}$ with the first NN
distance of the bulk gas, and $\lambda_r$ with one of the following
characteristic lengths: a) side length of the central pentagon in the
potential plots in fig. \ref{fig_potMINandHist} ($\lambda_s=0.45$nm), b)
distance between the center of the central pentagon and one of its
vertices ($\lambda_c=0.40$nm), c) $L=\tau\cdot S=0.45$ nm, where
$\tau$ = 1.618 is the golden ratio of the QC and $S$ = 0.243 nm is the
side length of the rhombic Penrose tiles \cite{qc1}.  Although there
is no {\it a priori} reason to choose one definition over the others,
the one that we select (equation $\ref{delta_m}$) has the
convenience of being perfectly correlated with the presence of the
transition, and of using reference lengths commonly determined in
experimental measurements ($\lambda_r$) or quantities
easy to extract ($k\cdot\sigma_{gg}$).

\begin{figure}
\begin{indented}
\item[]
\epsfig{file=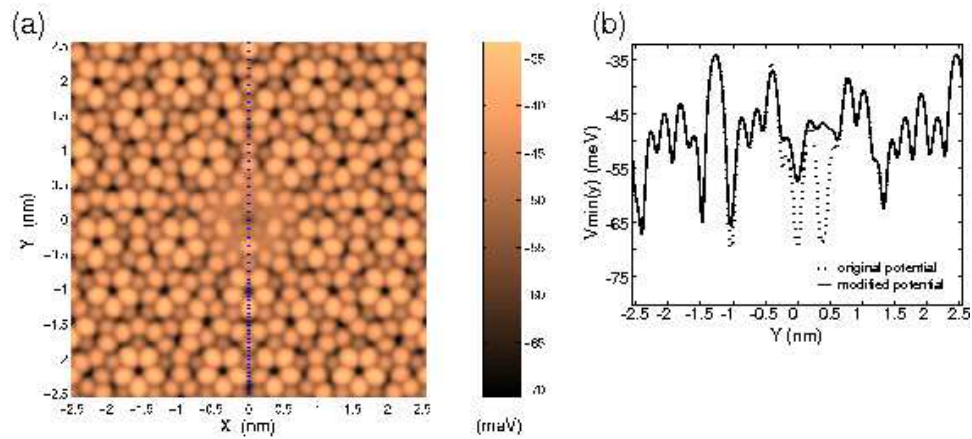,height=58mm,clip=}
 \vspace{1mm}
 \caption{\small
 (color online). 
   (a) The minimum of adsorption potential,
   $V_{min}(x,y)$, for Ne on a smoothed QC as described in the text. 
   (b) The variations of the minimum 
   adsorption potentials along the line at $x=0$ shown in (a), 
   for the modified and original interactions (solid and dotted curves).
 }
 \label{fig_potMINmod}
\end{indented}
\end{figure}

In figure \ref{fig_potMINandHist} we can observe that near the center of
each potential there is a set of five points with the highest binding
interaction (the dark spots constituting the central pentagons).  A
real QC surface contains an infinite number of these very attractive
positions which are located at regular distances and with five fold
symmetry.  Due to the limited size and shape of the simulation cell,
our surface contains only one set of these points.  Therefore, it is
of our concern to check if the results regarding the existence of the
transition are real or artifacts of the method.  We perform simulation
tests by mitigating the effect of the attractive spots through a
Gaussian smoothing function which reduces the corrugation of the
original potential.  The definitions are the following:
\begin{eqnarray}
  &&G(x,y,z) \equiv A_G e^{-(x^2+y^2+z^2)/2\sigma^2_G}, \\
  &&V(z) \equiv \left<V(x,y,z)\right>_{(x,y)},\\
  &&V_{mod}(x,y,z) \equiv V(x,y,z)\cdot \left[1-G(x,y,z)\right] + V(z)\cdot G(x,y,z).
\end{eqnarray}
where $G(x,y,z)$ is the Gaussian smoothing function (centered on the
origin and with parameters $A_G$ and $\sigma_G$), $V(z)$ is the
average over $(x,y)$ of the original potential $V(x,y,z)$, and
$V_{mod}(x,y,z)$ is the final smoothed interaction.  An example is
shown in figure \ref{fig_potMINmod}(a) where we plot the minimum of the
adsorption potential for a Ne/QC modified interaction (smoothed using
$A_G=0.5$ and $\sigma_G=0.4$ nm).  In addition, in panel (b) we show
the variations of the minimum adsorption potentials along line $x=0$
for the modified and original interactions (solid and dotted curves,
respectively).

Using the modified interactions (with $A_G=0.5$ and $\sigma_G=0.4$ nm)
we simulate all the noble gases of table \ref{table_mismatch}.  The
results regarding the phase transition on modified surfaces do not
differ from those on unmodified ones, confirming that the observed
transition behavior is a consequence of competing interactions between
the adsorbate and the whole QC substrate rather than just depinning of
the monolayer epitaxially nucleated.  Therefore, the simple criterion
for the existence of the transition ($\delta_m>0$) might also be
relevant for predicting such phenomena on other decagonal
quasicrystal substrates.

\section{Conclusions}
\label{section_conclusion}

We have presented the results of GCMC simulations of noble gas films
on QC. Ne, Ar, Kr, and Xe grow layer-by-layer at low temperatures up
to several layers before bulk condensation.  We observe interesting
phenomena that can only be attributed to the quasicrystallinity and/or
corrugation of the substrate, including structural evolution of the
overlayer films from commensurate pentagonal to incommensurate
triangular, substrate-induced alignment of the incommensurate films,
and density increase in each layer with the largest one observed in
the first layer and in the smallest gas.  Two-dimensional
quasicrystalline epitaxial structures of the overlayer form in all the
systems only in the monolayer regime and at low pressure.  The final
structure of the films is a triangular lattice with a considerable
amount of defects except in Xe/QC. Here a first-order transition
occurs in the monolayer regime resulting in an almost perfect
triangular lattice.  The subsequent layers of Xe/QC have hexagonal
close-packed structures.  By simulating fictitious systems with
various sizes and strengths, we find that the dimension of the noble
gas, $\sigma_{gg}$, is the most crucial parameter in determining the
existence of the phenomenon which is found only in systems with
$\sigma_{gg}\ge\sigma_{gg}$[Xe]. The results of this study will be
investigated in future experiments carried out in this laboratory,
analogous to those already performed with Xe.

This work stimulates more comprehensive studies of adsorption
heterogeneities on QC to understand the fundamental factors
controlling the structure of single-element films by elucidating
phenomena that are due exclusively to the quasiperiodicity of the
substrate, as opposed to chemical interactions between the adsorbates
and the surface.

\ack

We wish to acknowledge helpful discussions with 
L W Bruch, M Widom, A N Kolmogorov, C Henley, L Howle, D Rabson and R A Trasca.
We acknowledge the San Diego Supercomputer
Center for computing time under Proposal Number MSS060002.
This research was supported by NSF grant DMR-0505160.

\end{document}